\documentclass[twocolumn,showpacs,preprintnumbers,amsmath,amssymb,groupedaddress,pre]{revtex4}

\usepackage{graphicx}
\usepackage{dcolumn}
\usepackage{bm}
\usepackage{epsfig}

\def\e{\begin{equation}}
\def\f{\end{equation}}
\def\=#1{\overline{\overline #1}}

\def\-#1{{\bf #1}}
\def\.{\cdot}
\def\l#1{\label{eq:#1}}
\def\r#1{(\ref{eq:#1})}
\def\vec#1{{\bf #1}}

\begin{document}

\title{Subwavelength metallic waveguides loaded by uniaxial resonant scatterers}

\author{Pavel A. Belov}
\affiliation{Mobile Communication Division, Telecommunication Network Business, Samsung Electronics Co., Ltd., \\
94-1, Imsoo-Dong, Gumi-City, Gyeong-Buk, 730-350, Korea}

\email{belov@rain.ifmo.ru}

\author{Constantin R. Simovski}
\affiliation{Photonics and Optoinformatics Department, St.
Petersburg State University of Information Technologies, Mechanics
and Optics, Sablinskaya 14, 197101, St. Petersburg, Russia}

\date{\today}

\begin{abstract}
The dispersion properties of rectangular metallic waveguides
periodically loaded by uniaxial resonant scatterers are studied
with help of an analytical theory based on the local field
approach, the dipole approximation and the method of images. 
The cases of both magnetic and electric uniaxial scatterers with both longitudinal and
transverse orientations with respect to the waveguide axis are considered.
It is shown that in all considered cases waveguides support propagating modes 
below cutoff of the hollow waveguide within some frequency bands 
near the resonant frequency of the individual scatterers.
The modes are forward ones except the case of transversely oriented magnetic scatterers
when the mode turns out to be backward. The described effects can be applied for 
the miniaturization of the guiding structures.
\end{abstract}

\maketitle

\section{Introduction}

\begin{figure}[h]
\centering \epsfig{file=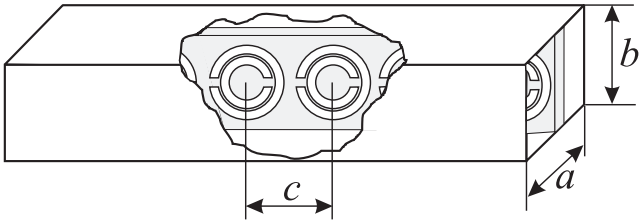, width=8.5cm}
\caption{Geometry of a subwavelength Split-Ring-Resonator-loaded
metallic waveguide} \label{geometry}
\end{figure}

Recently, a very unusual waveguide was proposed by R. Marques in \cite{Marqueswaveguide} and 
then extensively studied by S. Hrabar in \cite{Hrabarwaveguide}. It is a rectangular metallic waveguide
periodically loaded by resonant magnetic scatterers, so-called split-ring-resonators (SRR:s) \cite{PendrySRR,MarquesSRR},
which are also used as components of a realization of the left-handed medium (LHM) \cite{SmithWSRR}, composite
with negative permittivity and permeability \cite{Veselago,Metaspecial}. 
The geometry of the Marques waveguide (MW) is presented in Fig.\ref{geometry}.
The SRR:s in MW are oriented so that their magnetic moments are orthogonal to the waveguide axis and to one of
the walls. 

The MW support a propagating mode within a frequency band near the resonance of SRR:s even if it is located 
below cutoff frequency of the hollow waveguide \cite{Marqueswaveguide,Hrabarwaveguide}.
The transversal dimensions of the waveguide happen to be much smaller than the wavelength in free space. 
Thus, loading by SRR:s makes waveguide subwavelength and provide unique method for miniaturization of guiding structures.
The mode of MW is backward wave (the group velocity is negative). 
This effect was interpreted in \cite{Marqueswaveguide} in terms of the
effective LHM to which such the loaded waveguide is apparently equivalent.
The empty waveguide was considered as an artificial electric plasma with negative permittivity and the array of
magnetic scatterers as a magnetic material with negative permeability. 
This interpretation is not completely adequate because it cannot explain why the effect disappears in
the case of loading by isotropic magnetic with negative permeability. 
Really, it is clear that the hollow waveguide filled by 
isotropic magnetic with negative permeability does not support guiding modes. Meanwhile,
the doubly-negative medium having the isotropic negative permittivity and permeability 
would support propagating backward waves \cite{Lind}.
Also, the LHM interpretation is not instructive in our opinion since it does not allow to notice possibilities 
to obtain propagation below the cutoff frequency of hollow waveguide with help of the other loadings that SRR:s.

The goal of the present study is to give an adequate explanation for the extraordinary propagation effect in MW 
and to suggest another loadings which would lead to the similar effects.
In this paper it is shown that the propagation below the cutoff frequency of hollow waveguide can be achieved 
with magnetic scatterers oriented longitudinally with respect to waveguide axis, as well as with electric scatterers
oriented either longitudinally or transversally. This demonstrates that the miniaturization of the
rectangular waveguide at a fixed frequency using loading by the resonant scatterers is not restricted by the case 
when the scatterers are magnetic and transversally oriented.
The miniaturization is possible with help of either magnetic or electric resonant scatterers
with either transversal or longitudinal orientation with respect to the waveguide axis. 
Of course, the miniaturization can be also reached using periodically located capacitive posts, however the
loading by resonant scatterers is a qualitatively different effect.
In \cite{Hrabarwaveguide} it was pointed out that the miniaturization obtained in this way refers also to the
longitudinal size of the waveguide since the period of the loads
in incomparably smaller than the wavelength in free space, unlike
the propagation in a capacitively loaded waveguide, where the period of the posts is of the order of $\lambda/2$. 

The mini pass band below the cutoff frequency of a rectangular waveguide loaded by resonant scatterers is caused
by the properties of the periodical one-dimensional array (chain) of resonant dipoles and has nothing to do with doubly-negative
media. The backward wave appears in a special case of transverse orientation of magnetic scatterers and is not a necessary attribute of such mini-band. It is known that a chain of the resonant scatterers in a homogeneous
matrix supports guided modes. In the optical frequency range it refers to chains of metallic nanoparticles
\cite{Brongersma,Maier1,Maier2,Weber}. At microwaves it refers to so-called magneto-inductive waveguides (chains of SRR:s) \cite{Shamonina1,Shamonina2,Shamonina3,Shamonina4} or to chains of inductively loaded electric dipoles \cite{Tretlines}. 
The metallic walls of loaded waveguide perturb the dispersion of the
guided mode in a chain but do not cancel the propagation. This is because the
wavelength of the guided mode in the chain of resonant scatterers
is dramatically shortened compared to that in the matrix. As a
result, the energy of a guided mode is concentrated in a narrow
domain around the chain, and the interaction between the chain and
the waveguide walls turns out to be not critical for the existence
of the guided mode.

The paper is organized as follows. In the Section II the dispersion
properties of the chains of resonant scatterers located in free space are considered.
The known results, obtained in the work \cite{Weber} by
numerical simulations, are reproduced with help of analytical theory based on local field method. 
This is a necessary part of the work in the view of comparison with the case of the loaded waveguide.
The coincidence with known results can be considered as a validation of our approach. 
In the Section III the dispersion properties 
of the chains located inside of the rectangular waveguide are considered using two approaches:
an accurate method of local field (as in Section II) and an effective medium filling approximation. 
The Section IV is devoted to comparison between properties of the chains and loaded waveguides.
The Section V contains concluding remarks.The details of the local field theory are given in Appendix.

\begin{figure}[h]
\centering \epsfig{file=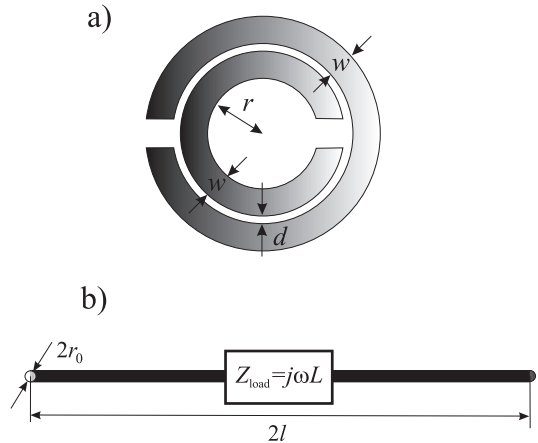, width=7cm} \caption{Geometries of
resonant scatterers: a) Split-Ring-Resonator, b) inductively loaded wire dipole.} \label{scat}
\end{figure}

In the present paper we consider both magnetic and electric
resonant uniaxial scatterers. As an example of a magnetic
scatterer we have chosen the SRR:s
\cite{PendrySRR,SmithWSRR,Shelbyscience} (see Fig. \ref{scat}.a). The
electric dipoles are represented in our work by the short
inductively loaded wires (ILW) \cite{LWD} (see Fig. \ref{scat}.b). Any
individual scatterer can be characterized by polarizability
relating the dipole moment (magnetic or electric) with
the local field (magnetic or electric external field acting to the
scatterer). This polarizability is scalar since the only possible
direction of the induced dipole moment is possible for a scatterer
with given orientation. The details concerning calculation of the polarizabilities
for SRR:s and ILW:s are presented in Appendix A.

The inverse values of the polarizabilities $\alpha(\omega)$ and $\alpha_e(\omega)$
(see Appendix A, formulae \r{alpha} and \r{alpe}) of SRR:s and ILW:s 
have the same dependencies on frequency within the resonant band: 
\e
{\rm Re} \{\alpha^{-1}(\omega)\}=A^{-1}\left(\frac{\omega_0^2}{\omega^2}-1\right).
\l{inva} 
\f 
Here $A$ is amplitude and $\omega_0$ is resonant frequency, the parameters determined
by the geometry of the scatterer. Notice, that the result \r{inva} is
also valid for a silver nanosphere in the vicinity of its plasmon
resonance \cite{Weber}. Thus, it is clear that there is no
principal difference between the dispersion properties of the
chain of SRR:s and ILW:s (at microwaves) or silver nanospheres (in the optical range).

\section{Chains of uniaxial resonant scatterers}

\begin{figure}[h]
\centering \epsfig{file=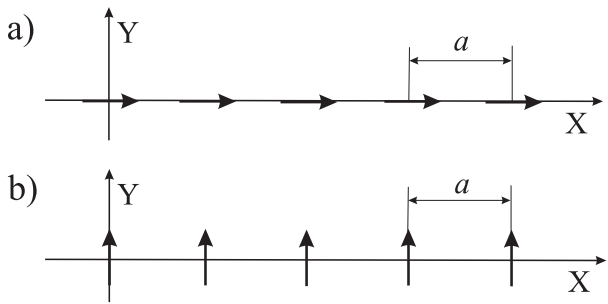, width=8cm}
\caption{Chains of resonant scatterers. a) Longitudinal orientation. b) Transverse orientation.}
\label{chains}
\end{figure}

Let us study dispersion properties of the linear chains with period
$a$ formed by resonant scatterers. We will consider only two
typical orientations of scatterers: longitudinal and transverse.
The geometries of the structures are presented in Fig. \ref{chains}. 
The case of longitudinal orientation was analyzed in \cite{Tretlines}, and the both 
longitudinal and transverse orientations were studied in \cite{Weber}. 
In the present section we reproduce the main results of these works with help of the local field approach.

\subsection{Basic theory}

Without loss of generality we can restrict consideration by the case of the chain of magnetic scatterers (SRR:s).
The chains of electric scatterers are dual structures to the considered ones and 
have completely the same dispersion properties.

The spatial distribution of dipole moments of SRR:s corresponding to an eigenmode of a chain 
is determined by a propagation constant $q$: $M_n=M e^{-jqan}$.
Following to the local field approach the dipole moment $M$ of a
reference (zeroth) scatterer can be expressed in terms of the magnetic field
$\-H_{\rm loc.}$ acting to it: $M=\alpha H^d_{\rm loc.}$, where
$H^d_{\rm loc.}=(\-H_{\rm loc.}\.\-d)$ is the projection of the
field on the direction of the scatterer ($\vec d=\vec x_0$ for longitudinal
orientation of scatterers and $\vec d=\vec y_0$ for transverse one). 
This local field is a sum of partial magnetic fields
$\-H_{m}$ produced at the coordinate origin by all other
scatterers with indexes $m\ne 0$: $ \-H_{\rm loc.}=\sum\limits_{m
\ne 0} \-H_{m}$.

The magnetic field produced by a single scatterer with dipole moment $\-M_m$
at a point with radius vector $\vec R$ is given by dyadic Green's function $\=G(\-R)$:
\e
\-H_m(\vec R)=\mu_0^{-1}\=G(\-R)\-M_m,
\l{Ggen}
\f
where
\e
\=G(\-R)=\left(k^2\=I+\nabla\nabla\right)\frac{e^{-jkR}}{4\pi R}.
\l{G}
\f

Since all dipole moments of the chain are oriented along $\-d$ it
is enough to use only the $\-d\-d$ component of dyadic Green's
function. So, we replace \r{Ggen} by the scalar expression: \e
H_m^d(\vec R)=\mu_0^{-1}G_{dd}(\-R)M_m, \l{Green} \f where \e
G_{dd}(\-R)=\left(k^2+\frac{\partial^2}{\partial
d^2}\right)\frac{e^{-jkR}}{4\pi R}, \l{gre}\f and $d$ means $x$
for the longitudinal case and $y$ for the transverse case.

Finally we obtain the expression for the field acting to the
reference scatterer in the form: \e H^d_{\rm
loc.}=\sum\limits_{m\ne 0} G_{dd}(am\vec x_0) e^{-jqam} M. \f

It allows to get dispersion equation for the chains under
consideration: 
\e 
\mu_0\alpha(\omega)^{-1}=C_{d}(\omega,q,a),
\l{displine} \f where
$$
C_{d}(\omega,q,a)=\sum\limits_{m\ne 0} G_{dd}(am\vec x_0) e^{-jqam}.
$$
In the Appendix B we provide expressions \r{Cx} and \r{Cy} 
which we use for effective numerical calculations of interaction constants $C_x$ and $C_y$,
corresponding to transverse and longitudinal orientations of scatterers, respectively.

\subsection{Analysis of dispersion properties}

The dispersion diagram for guided modes can be obtained solving
transcendental dispersion equation \r{displine} with interaction
constants given by expressions \r{Cx} and \r{Cy}. Geometrically,
dispersion curves correspond to the lines of level where the
surface plot of function ${\rm Re} \{C_{x,y}(\omega,q)\}$ is
crossed by $\mu_0 {\rm Re}\{\alpha^{-1}(\omega)\}$. Note, that
only the solutions with $k<q<2\pi/a-k$ for $k<\pi/a$ correspond to
guided modes: For $|q|<k$, ${\rm Im} \{ C_{x,y}(\omega,q)-
\mu_0\alpha^{-1}\} \ne 0$ and dispersion equation \r{displine} has
complex solutions corresponding to leaky modes (see details in Appendix B).

\begin{figure}[h]
\centering \epsfig{file=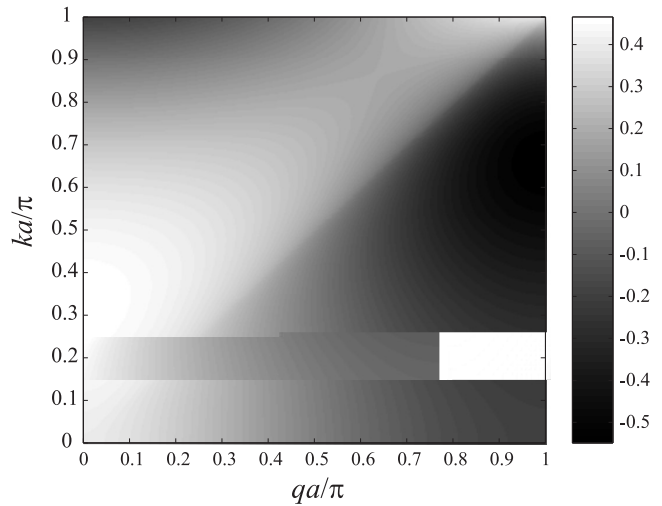, width=8.5cm}
\caption{Dependence of ${\rm Re}\{C_x a^3\}$ on normalized frequency $ka/\pi$ and propagation factor $qa/\pi$}
\label{cx}
\end{figure}
\begin{figure}[h]
\centering \epsfig{file=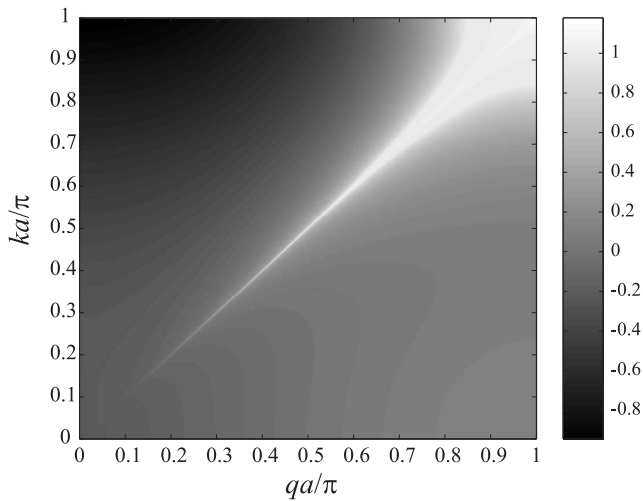, width=8.5cm}
\caption{Dependence of ${\rm Re}\{C_y a^3\}$ on normalized frequency $ka/\pi$ and propagation factor $qa/\pi$}
\label{cy}
\end{figure}

The dependencies of $C_x$ and $C_y$ on on normalized frequency
$ka/\pi$ and propagation factor $qa/\pi$ are shown in Fig.
\ref{cx} and \ref{cy}, respectively. The interaction constants 
vary in $[-1,1]$ range except the case of $C_y$ with $q$ close
to $k$, which has logarithmic singularity at the light line $q=k$.
The function $\mu_0 {\rm Re}\{\alpha^{-1}(\omega)\}$ decreases
very rapidly within $[-1,1]$ range of values near the resonant
frequency $\omega_0$. It means, that guided modes exist only
within a narrow bands near the resonant frequency of scatterers.
The behavior of dispersion curves can be easily predicted from the
plots in Fig. \ref{cx} and \ref{cy}. If at a fixed frequency the
interaction constant decays when propagation factor increases then
the dispersion curve grows and the eigenmode is forward wave (the group velocity $v_g=\frac{d\omega}{dq}$ is positive),
but if the interaction constant grows then the dispersion curve decays
and the eigenmode is backward wave (the group velocity $v_g=\frac{d\omega}{dq}$ is negative).
From Fig. \ref{cx} it is clear
that for any resonant frequencies satisfying to the evident
condition $\omega_0<\pi/(a\sqrt{\varepsilon_0\mu_0})$ 
(corresponding to the propagation below cutoff of the hollow waveguide )
the longitudinal mode is forward wave because $C_x$ decays versus $q$.
In the case of transverse modes the situation is different. 
While $k_0a<0.5\pi$ ($k_0=\omega_0\sqrt{\varepsilon_0\mu_0}$) a two mode regime holds.
The interaction constant $C_y$ decays while $q$ is close to $k$,
but from a certain $q$ it starts to grow. It means that one of the
transverse eigenmodes is forward (with $q\approx k $) and the other one is
backward. If the resonant frequency is high enough
($0.5\pi<k_0a<\pi$) the two mode regime disappears and only the
forward wave remains.

\begin{figure}[h]
\centering \epsfig{file=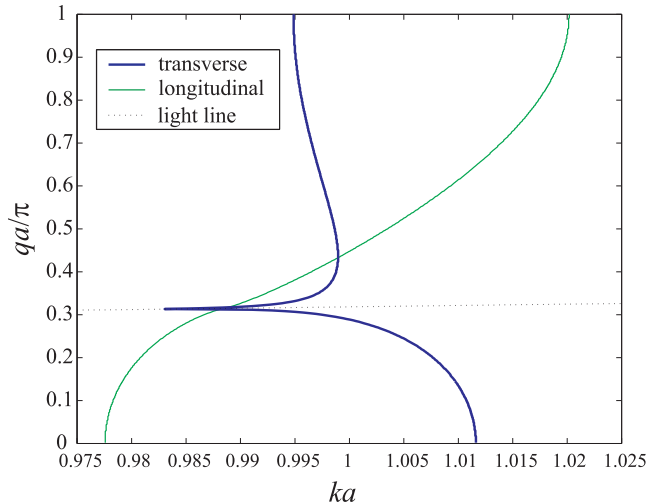, width=8.5cm}
\caption{Dispersion diagram for chains of resonant scatterers: transverse (thick line) and longitudinal (thin line)
orientations}
\label{displine}
\end{figure}

The typical dispersion diagram for both longitudinal and
transverse modes is presented in Fig. \ref{displine} for the case
of scatterers with $A=0.1\mu_0a^3$ and $\omega_0
a=1/\sqrt{\varepsilon_0\mu_0}$. The similar dispersion diagram was
obtained in \cite{Weber} (see Fig. 3 of this work) by a numerical
simulation. Though in \cite{Weber} the electric scatterers in the optical range were considered, 
but we consider the magnetic scatterers in the microwave range, the use of duality principle, and normalized frequency
$ka/\pi$ and wave vector $qa/\pi$ eliminates this difference. 
The polarizability of silver nanospheres, for which Fig. 3 from \cite{Weber} was obtained, obeys to
expression (8) of \cite{Weber} which is identical to our formula \r{inva}).

As it was predicted, the longitudinal mode is forward wave and there is
a two mode regime for transverse modes. The dispersion curve for
transverse waves has the asymptote $q=k$ and both leaky ($q<k$) and guided ($q>k$)
modes exist at very low frequencies where they have almost equal
wave vectors. This fact indicates the good matching between the
radiated wave and the guided mode. Within the band $0.995<ka<1$
there are two guiding modes at every frequency. The solution corresponding
to the backward wave is close to the Bragg's mode ($qa\approx
\pi$) whose group velocity is close to zero. The field of this
mode is concentrated near the chain within the spatial region
$r=\sqrt{z^2+y^2}<1/\sqrt{q^2-k^2}\sim a$. The same concerns the
longitudinal mode within the band $1.015<ka<1.020$. If the period
of the chain is much smaller than wavelength in free space the
waveguide is sub-wavelength (the field of the guided mode is
concentrated within a cylindrical domain whose diameter is much
smaller than $\lambda$). Thus, the chains of resonant scatterers
(electric or magnetic, it does not matter) form sub-wavelength
waveguides which can support either forward or backward waves
\cite{Brongersma,Maier1,Maier2,Weber,Tretlines,Shamonina1,Shamonina2,Shamonina3,Shamonina4}.

\section{Loaded waveguides}

\subsection{Basic theory}

Let us study the eigenmodes of the rectangular metallic waveguides
periodically loaded by resonant uniaxial scatterers. Such
structures can be effectively considered as linear chains located
inside of the metallic waveguides. The geometries of the four
waveguides considered in the present paper are shown in Fig.
\ref{geometry} (left sides of subplots). The chains with period
$c$ along the waveguide axis are located at the center of rectangular metallic waveguides with
dimensions $a\times b$. The structures in Figs. \ref{geometry}.a-d differ by orientation of
scatterers (longitudinal or transverse) and their type (electric or magnetic). 
The first structure (with transversely oriented magnetic scatterers) is the sub-wavelength waveguide (see Fig.
\ref{geometry}) suggested by R. Marques \cite{Marqueswaveguide,Hrabarwaveguide}. The other ones are
considered in order to show three other possible ways to obtain miniaturization of rectangular waveguides.

Note, that the chains of electric scatterers are no more dual to the chains of magnetic scatterers 
(in contrast to the chains in free space)
due to the different interaction of electric and magnetic dipoles with metallic walls.

\begin{figure}[h]
\centering \epsfig{file=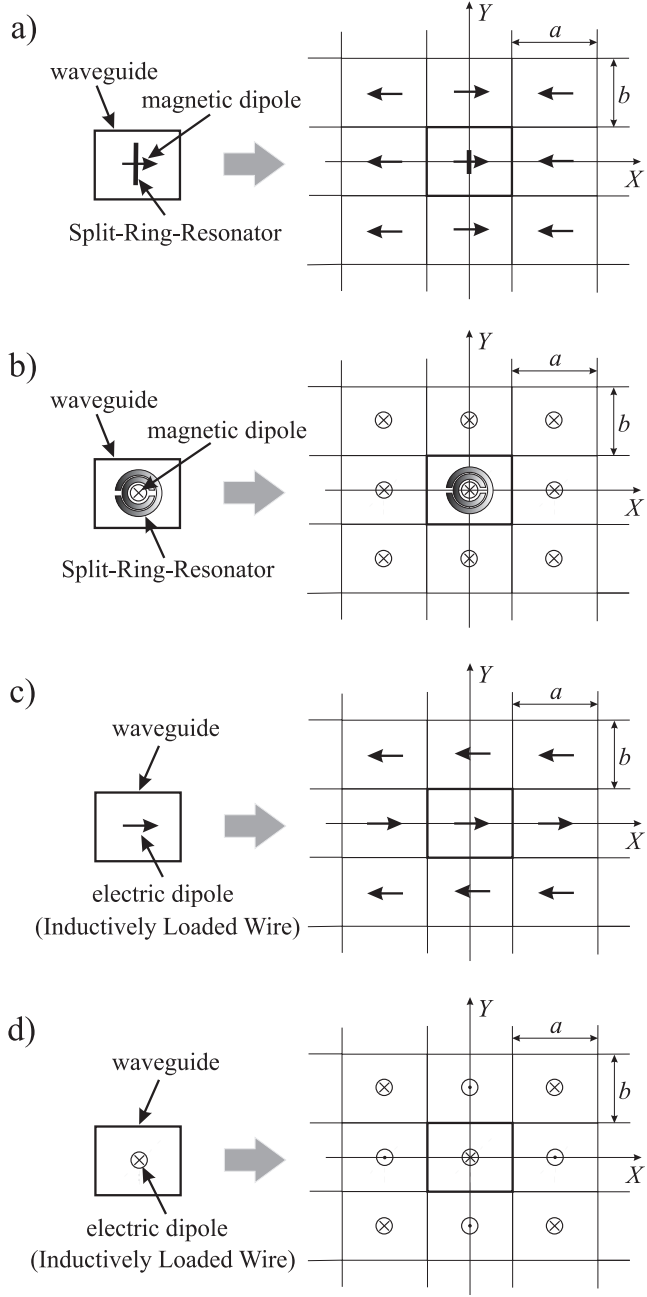, width=8.5cm}
\caption{Transformation of the waveguide problem to the lattice
one with the use of the image principle} \label{image}
\end{figure}

The dispersion equation for the chains keeps the same form as \r{displine}, but
free-space dyadic Green's function $\=G(\vec R)$ \r{G} should be replaced by Green's
function of the waveguide which takes into account the metallic walls.
This Green's function can be determined with help of the image principle. 
This approach transforms the eigenmode problem for the loaded waveguide 
to the problem of the eigenwave propagation in a three-dimensional
electromagnetic lattice formed by same scatterers. The details of
the transformation are illustrated by Fig. \ref{image} (right
parts of all subplots). The electromagnetic crystals obtained in
such a way have orthorhombic elementary cell $a\times b\times c$
and their dispersion properties was studied in \cite{Belovhomo}
using local field approach. Thus, we can apply the theory of the
electromagnetic interaction in dipole crystals presented in
\cite{Belovhomo} in order to study dispersion properties of 
waveguides under consideration.

In the coordinate system associated to the axes of the crystal
the center of a scatterer with indexes $(m,n,l)$ has coordinates $\vec R_{m,n,l}=(am,bn,cl)^T$. 
From Fig. \ref{image} it is clear, that distribution of dipole moments 
in the lattice has the following form:
$$
\vec M_{m,n,l}=(-1)^mMe^{-jqcl}\vec x_0,
$$
for the case of transverse orientation of magnetic scatterers (Fig. \ref{image}.a);
$$
\vec M_{m,n,l}=Me^{-jqcl}\vec z_0,
$$
for the case of longitudinal orientation of magnetic scatterers (Fig. \ref{image}.b);
$$
\vec P_{m,n,l}=(-1)^nPe^{-jqcl}\vec x_0,
$$
for the case of transverse orientation of electric scatterers (Fig. \ref{image}.c); and
$$
\vec P_{m,n,l}=(-1)^{m+n}Pe^{-jqcl}\vec z_0,
$$
for the case of longitudinal orientation of electric scatterers (Fig. \ref{image}.d).

Any of these distributions can be rewritten in terms of a
wavevector $\vec q$ as $e^{-j(\vec q\. \vec R_{m,n,l})}$, where
the wavevector $\vec q$ for four cases considered above has the
form $(\pi/a,0,q)^T$, $(0,0,q)^T$, $(0,\pi/b,q)^T$ and
$(\pi/a,\pi/b,q)^T$, respectively. This notation finally makes
clear that the waveguide dispersion problems reduce to those of the
three-dimensional lattices in the special cases of certain propagation directions.

The dispersion equation for three-dimensional electromagnetic
crystal formed by magnetic scatterers oriented along $x$-axis has the form \cite{Belovhomo}:
\e 
\mu_0\alpha^{-1}(\omega)-C(k,\-q)=0, \l{disper} 
\f
where \e C(k,\-q,a,b,c)=\sum\limits_{(m,n,l)\ne(0,0,0)}
G(\-R_{m,n,l}) e^{-j(q_xam+q_ybn+q_zcl)}. \l{C} \f
We call $C(k,\-q,a,b,c)$ as the dynamic interaction constant of
the lattice using the analogy with the classical interaction
constant from the theory of artificial dielectrics and magnetics
\cite{Collin}. The explicit expression for $C$ for the general
case was derived in \cite{Belovhomo} and it is given in Appendix C by formula \r{cfinal}.

The dispersion equation for waveguides with transverse orientation of scatterers
can be directly obtained from equation \r{disper} 
by substitution $\vec q=(\pi/a,0,q)^T$ and $\vec q=(0,\pi/b,q)^T$ for magnetic and electric scatterers, respectively.
Also, in the case of electric scatterers following duality principle $\mu_0\alpha^{-1}(\omega)$
should be replaced by $\varepsilon_0\alpha_e^{-1}(\omega)$.
The similar operation for case of longitudinal orientation happens to be possible only after
rotation of coordinate axes: $z \rightarrow x'$, $x \rightarrow y'$, $y \rightarrow z'$, since equation \r{disper} requires scatterers to be directed along $x$-axis, but for longitudinal orientation they are directed along $z$-axis.
After such manipulation substitution of $\vec q=(q,0,0)^T$ and $\vec q=(q,\pi/a,\pi/b)^T$ 
(in the new coordinate axes $(x',y',z')$) into equation \r{disper} provide desired dispersion equations
for waveguides with transverse orientation of magnetic and electric scatterers, respectively.

This way we obtain the following dispersion equations for all loaded waveguides under consideration: \e
\mu_0\alpha^{-1}(\omega)-C(k,(\pi/a,0,q)^T,a,b,c)=0, \l{dispmt} \f
for transverse orientation of magnetic scatterers; \e
\mu_0\alpha^{-1}(\omega)-C(k,(q,0,0)^T,c,a,b)=0, \l{dispml} \f for
longitudinal orientation of magnetic scatterers; \e
\varepsilon_0\alpha_e^{-1}(\omega)-C(k,(0,\pi/b,q)^T,a,b,c)=0,
\l{disppt} \f for transverse orientation of electric scatterers;
and \e
\varepsilon_0\alpha_e^{-1}(\omega)-C(k,(q,\pi/a,\pi/b)^T,c,a,b)=0,
\l{disppl} \f for longitudinal orientation of electric scatterers.

The obtained dispersion equations are real valued ones, because the imaginary parts of
its components cancels out. It can be clearly seen from Sipe-Kronendonk condition \r{sipe}
and following expression proved in \cite{Belovhomo}:
\e
{\rm Im}(C)=\frac{k^3}{6\pi}.
\l{imc} \f

\subsection{Effective medium filling approximation}

The chain of SRR:s with transverse orientation located in the waveguide has
been interpreted in the literature as a piece of an uniaxial magnetic medium \cite{Marqueswaveguide,Hrabarwaveguide}. 
We call this approach as effective medium filling approximation.
It can be applied practically to every waveguide considered in this paper except the case 
of longitudinal orientation of magnetic scatterers 
since the uniaxial magnetic model does not describe longitudinal modes.
This approach provide qualitatively acceptable results which are compared with exact ones in the next subsection.

Let us start from the case of transversely oriented magnetic
scatterers (Fig. \ref{image}.a) and consider a chain of parallel uniaxial magnetic
scatterers as a piece of infinite resonant uniaxial magnetic. The
permeability of such a magnetic is a tensor (dyadic) of the form:
$$
\=\mu=\mu \-x_0\-x_0+\mu_0(\-y_0\-y_0+\-z_0\-z_0).
$$

The permeability $\mu$ along the anisotropy axis $x$, can be
calculated though the individual polarizability of a single
scatterer using the Clausius-Mossotti formula: 
\e
\mu=\mu_0\left(1+\frac{\alpha(\omega)/(\mu_0V)}{1-C_s(a,b,c)
\alpha (\omega)/\mu_0}\right), \l{CM} 
\f where $V=abc$ is a volume of the elementary cell of an infinite three-dimensional lattice 
and $C_s(a,b,c)$ is the known static interaction constant of the lattice
\cite{Collin,Belovhomo}. In the case of a simple cubic lattice
$a=b=c$ the interaction constant is equal to the classical value
$C_s=1/(3V)$.

Notice, that we should skip the radiation losses contribution in
expression \r{invalph} while substituting into formula \r{CM}.
This makes permeability purely real number as it should be for
lossless material. This manipulation is based on the fact that the
far-field radiation of the single scatterer is compensated by the
electromagnetic interaction in a regular three-dimensional array,
so that there were no radiation losses for the wave propagating in
the lattice \cite{Sipe,Belovhomo}.

The dispersion equation for the uniaxial magnetic medium has the
following form (see e.g. \cite{BelovMOTL}): 
\e 
\mu_0 (q_y^2+q_z^2)=\mu (k^2-q_x^2). \l{dispunis} 
\f 
To solve the waveguide dispersion problem is to solve the dispersion problem
\r{dispunis} for a special case $\vec q=(\pi/a,0,q)^T$. The
substitution of $\vec q=(\pi/a,0,q)^T$ into \r{dispunis} gives: 
\e
q^2=\frac{\mu}{\mu_0}\left[k^2-\left(\frac{\pi}{a}\right)^2\right].
\l{qq} \f

For the frequencies below cutoff of the hollow waveguide the expression in the brackets of \r{qq} is
negative and for positive $\mu$ there is no propagation. However,
if $\mu$ is negative (it happens in accordance with \r{alpha} and
\r{CM} within a narrow frequency range near the resonance of the
scatterers, just above the resonant frequency of the media) $q$
becomes real. This mini pass-band can be located much lower than the
cutoff frequency of the empty waveguide with help of reduction of resonant frequency of the scatterers.
It is easy to see from \r{qq} that the mode is backward wave, i.e. $\frac{dq}{d\omega}<0$.
This follows from the basic inequality $\frac{d\mu}{d\omega}>0$ (Foster's theorem).

So, for transverse magnetic scatterers the effective medium filling model 
gives (at least qualitatively) a correct result. Namely, it predicts a mini-band within the resonance
band of SRRs and the backward wave propagating within it. 
However, this model is not accurate. The reason of this inaccuracy is simple. 
In spite of the low frequency of operation 
(the waveguide dimensions are small compared to the wavelength {\it in
free space}) the effective magnetic medium should operate in the
regime when its period is comparable with the wavelength {\it in
the effective medium} because $q_x=\pi/a$. Such regime for an electromagnetic
crystal can not be described with help of homogenization and requires
taking into account spatial resonances of the lattice \cite{Belovhomo}.

In the case of transversely oriented electric scatterers (Fig. \ref{image}.c) the
effective medium filling model implies that the wave propagates in a
uniaxial dielectric with permittivity tensor:
$$
\=\varepsilon=\mu \-x_0\-x_0+\varepsilon_0(\-y_0\-y_0+\-z_0\-z_0).
$$

The permittivity $\varepsilon$ along the anisotropy axis $x$ is
given by the Clausius-Mossotti formula: \e
\varepsilon=\varepsilon_0\left(1+\frac{\alpha_e(\omega)/(\varepsilon_0V)}{1-C_s(a,b,c)
\alpha_e (\omega)/\varepsilon_0}\right). \l{CMe} \f 
The dispersion equation for such uniaxial dielectric reads \cite{BelovMOTL}: 
\e
\varepsilon_0 (q_y^2+q_z^2)=\varepsilon (k^2-q_x^2). \l{dispunise}
\f 
Solution of waveguide dispersion problem corresponds to the
case when $\vec q=(0,\pi/b,q)^T$: \e
q^2=\frac{\varepsilon}{\varepsilon_0}k^2-\left(\frac{\pi}{b}\right)^2.
\l{qqe} 
\f 
This mode propagates only at the frequencies when permittivity takes high positive values
$\varepsilon>\varepsilon_0[\pi/(kb)]^2$. In our case of resonant
dielectric it happens within a mini-band just {\it below} the resonance of the medium.
It is clear from \r{qqe} that the mode is forward wave, i.e.
$\frac{dq}{d\omega}>0$. This follows from the Foster's theorem $\frac{d\varepsilon}{d\omega}>0$.

In the case of longitudinally oriented electric scatterers (Fig.
\ref{image}.d) the solution of the waveguide dispersion problem
can be obtained from the dispersion equation \r{dispunise} with
$\vec q=(q,\pi/a,\pi/b)^T$ (the axis were transformed in order to
have x-axis along dipoles) in the next form: \e
q^2=k^2-\frac{\varepsilon_0}{\varepsilon}
\left[\left(\frac{\pi}{a}\right)^2 +
\left(\frac{\pi}{b}\right)^2\right]. \l{qqe2} \f 
The mode is propagating at the frequencies when permittivity is positive and rather high, or negative.
This mode is forward wave in the same manner as \r{qqe} since $\frac{d\varepsilon}{d\omega}>0$.

\subsection{Analysis of dispersion properties}

For numerical calculation of dispersion curves using \r{dispmt}, \r{dispml},
\r{disppt}, \r{disppl} and  \r{cfinal} we have chosen square
waveguides ($a=b=c$) loaded by scatterers with the same parameters
which were used for studies of chains: 
$\omega_0=1/(a\sqrt{\varepsilon_0\mu_0})$, and $A=0.1\mu_0a^3$ for magnetic scatterers,
and $A_e=0.1\epsilon_0a^3$ for electric ones.

\begin{figure}[h]
\centering \epsfig{file=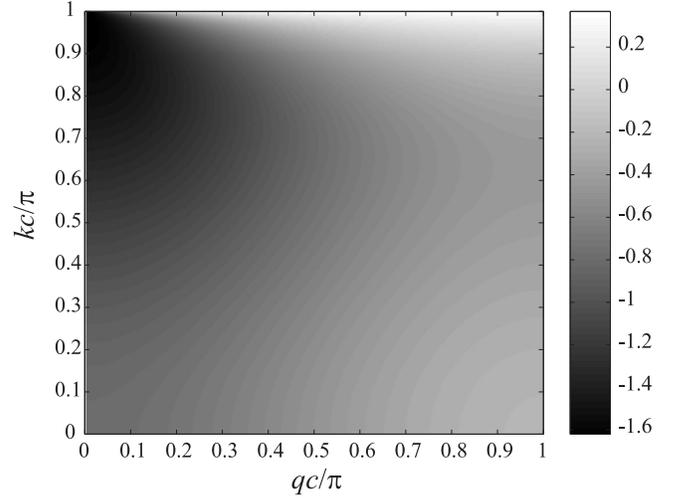, width=8.5cm}
\caption{Dependence of the real part of normalized interaction
constant $C(k,\vec q)a^3$ with $\vec q=(\pi/a,0,q)^T$
(corresponding to transverse orientation of magnetic scatterers)
 on normalized frequency $ka/\pi$ and propagation constant $qa/\pi$.
} \label{surfm01}
\end{figure}

The dependence of the real part of normalized interaction constant
$C(k,\vec q)$ with $\vec q=(\pi/a,0,q)^T$ on the normalized frequency
$ka/\pi$ and on the propagation constant $qa/\pi$ is presented in
Fig. \ref{surfm01}. This interaction constant corresponds to dispersion equation
\r{dispmt} for transverse orientation of magnetic scatterers. The
value of $\mbox{Re} (C)a^3$ varies within $[-2,0.5]$ interval
while the normalized frequency $ka/\pi$ is bounded by unity (which
corresponds to the cutoff of hollow waveguide). If a value of the normalized frequency
is fixed then the real part of the interaction constant is a monotonously growing function of $qa/\pi$.
The dependence of the real part of the interaction constant on
frequency is quite weak as compared with rapidly decreasing
$\alpha^{-1}(\omega)$ as follows from \r{invalph}. The function
$\mu_0 \alpha^{-1}(\omega)$ takes values within $[-2,0.5]$
interval at the frequencies close to resonant frequency $\omega_0$.
Therefore, dispersion equation \r{dispmt} has a real solution for $qa/\pi$
within a mini-band of frequencies near the resonant frequency of
the scatterers $\omega_0$, and this solution is a decaying function of frequency 
which corresponds to the backward wave (the group velocity $v_g=\frac{d\omega}{dq}$ is negative).
The obtained result, of course, confirms existence of backward wave below cutoff
of the hollow waveguide predicted and experimentally demonstrated in \cite{Marqueswaveguide,Hrabarwaveguide}.

\begin{figure}[h]
\centering \epsfig{file=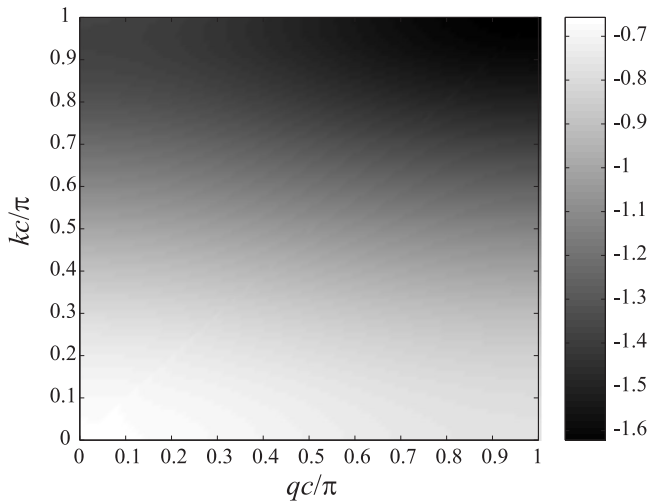, width=8.5cm}
\caption{Dependence of the real part of normalized interaction
constant $C(k,\vec q)a^3$ with $\vec q=(q,0,0)^T$
(corresponding to longitudinal orientation of magnetic scatterers)
 on normalized frequency $ka/\pi$ and propagation constant $qa/\pi$.
} \label{surfm00}
\end{figure}

In contrast to Fig. \ref{surfm01}, the dependencies of normalized
interaction constants $C(k,\vec q)$ with $\vec q=(q,0,0)^T$, $\vec
q=(0,\pi/a,q)^T$ and $\vec q=(q,\pi/a,\pi/a)^T$ are decaying
functions of $q$ for the fixed values of $k<\pi/a$. It means that
solutions of dispersion equations \r{dispml}, \r{disppt} and
\r{disppl} are forward waves for any resonant frequency of the
scatterer below $\pi/(a\sqrt{\varepsilon_0\mu_0})$.
The Fig. \ref{surfm00} shows dependence of the real part of normalized interaction constant
$C(k,\vec q)$ with $\vec q=(q,0,0)^T$ on the normalized frequency
$ka/\pi$ and on the propagation constant $qa/\pi$.
The dependencies for the cases $\vec q=(0,\pi/a,q)^T$ and $\vec q=(q,\pi/a,\pi/a)^T$
are not shown in order to reduce size of the paper.

\begin{figure}[h]
\centering \epsfig{file=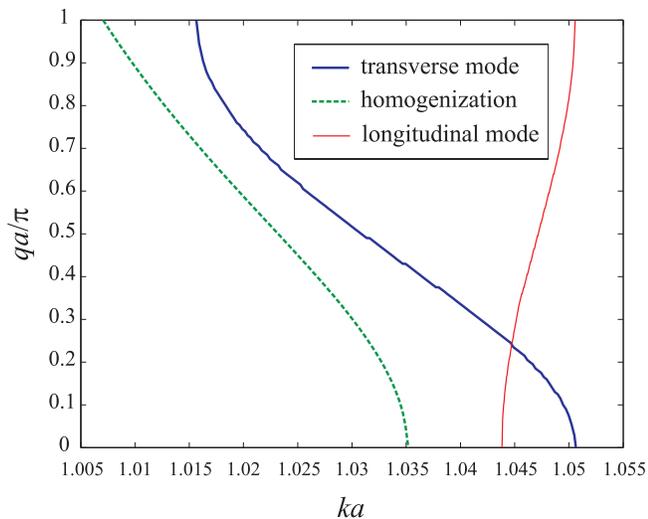, width=8.5cm}
\caption{Dispersion curves for metallic waveguides loaded by
magnetic scatterers: exact solution (thick line) and
effective medium filling approximation (dashed line) for transverse
orientation, and exact solution (thin line) for longitudinal
orientation. The effective medium filling model for the longitudinal case is
not applicable.} \label{dispm01}
\end{figure}

The dispersion curves for the case of magnetic scatterers are
presented in Fig. \ref{dispm01}. The thick solid line represents
the dispersion curve for the transverse mode. It is obtained by
numerical solution of transcendental dispersion equation
\r{dispmt}. The dashed line shows the result predicted by the
model of effective medium filling \r{qq}. The significant frequency shift
between the exact and approximate solutions is observed. Also, the
effective medium model gives wittingly wrong results with
$q>\pi/a$ in the region $ka<1.0055$, and incorrectly describes
group velocity for $q>\pi/(2a)$ (for example,
it does not describe the Bragg mode with zero group velocity at
the point $qa=\pi$). The dispersion curve for the
longitudinal mode obtained by numerical solution of equation \r{dispml}
is represented by thin line in Fig. \ref{dispm01}. 
As it was mentioned above, the effective medium filling model
can not be applied for description of this mode.

\begin{figure}[h]
\centering \epsfig{file=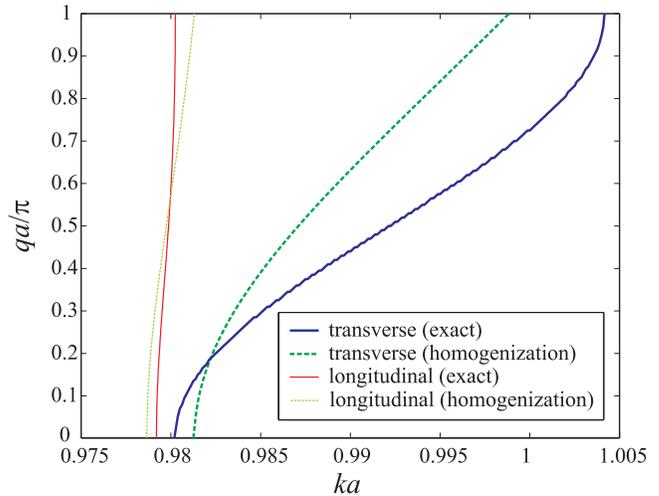, width=8.5cm}
\caption{Dispersion curves for metallic waveguides loaded by
electric scatterers: exact solution (thick line) and
effective medium filling approximation (dashed line) for transverse
orientation, and exact solution (thin line) and effective medium filling
approximation (thin dashed line) for longitudinal orientation. }
\label{dispp01}
\end{figure}
The dispersion curves for the case of electric scatterers are
presented in Fig. \ref{dispp01}. The thick and thin solid lines shows dispersion curves
for transverse and longitudinal modes, obtained by numerical solution
of dispersion equations \r{disppt}  and \r{disppl}, respectively. 
The dispersion curves provided by effective medium models for transverse and longitudinal modes 
(formulae \r{qqe} and  \r{qqe2})  are plotted by thick and thin dashed lines, respectively. 
The comparison of exact and approximate solutions shows that the model of effective medium
filling gives qualitatively right prediction of dispersion curves behavior in the case of electric scatterers
as well as in the case of transverse magnetic scatterers.
The drawbacks are also the same: wrong group velocity $q>\pi/(2a)$ and wittingly wrong results with
$q>\pi/a$ at some frequencies.

The Figs. \ref{dispm01} and \ref{dispp01} demonstrate that
the waveguides loaded by electric and magnetic resonant scatterers support
modes within the mini-bands below the cutoff frequency of the hollow waveguide.
The modes are forward waves, except the case of transverse magnetic scatterers
when the mode is forward wave. The bandwidth in the case of transverse electric scatterers
is of the same order with bandwidth for magnetic scatterers with the same parameters obtained with help of duality principle,
but the bandwidths for the longitudinal modes are significantly narrower than for the transverse ones.

\section{Discussion}

In the papers \cite{Marqueswaveguide,Hrabarwaveguide} the term `subwavelength waveguide' was applied
to a rectangular waveguide with small transversal dimensions as
compared to the wavelength in free space. However, there is a lot
of other works \cite{Brongersma,Maier1,Maier2,Weber,Tretlines,Shamonina1,Shamonina2,Shamonina3,Shamonina4}. 
in which the term `subwavelength waveguiding' means the
propagation of a wave along the chain of electrically small
nearly-resonant particles below the diffraction limit.
In this case the transversal size of the spatial domain, in which the field of the guided mode is concentrated, 
is much smaller that the wavelength in free space. Therefore, this mechanism of the wave transmission
is considered as prospective for subwavelength imaging.

Since the field of the mode guided along the chain of resonant
scatterers is concentrated within a subwavelength cross section,
the presence of the metal walls even at the rather small distance
turns out to be not crucial for the existence of the guided wave.
Thus, any waveguide periodically loaded by the scatterers can be considered as
a subwavelength waveguide formed by the chain of nearly resonant scatterers, whose
dispersive properties are perturbed by the metal walls. These
walls can be described in terms of the image chains forming an
infinite lattice. However, the wave propagates along the same
direction in every image chain, and the transversal wave
numbers $q_x=\pi/a$ or $q_y=\pi/b$ describe the transversal phase
distribution of the wave propagating along $z$ but not the energy
transport across $z$. The main question is how the image chains of
scatterers interact with the real chain, and how this interaction
influences its dispersive properties.

For that purpose let us compare  Fig. \ref{displine} with Figs. \ref{dispm01}
and \ref{dispp01}. We can conclude that the presence of metal
walls around the chain of resonant scatterers produces the
following effects: 
\begin{itemize}
\item It decreases the group velocity and the frequency band of the
guided mode which corresponds to the longitudinal orientation of
dipoles. 
\item It cancels the two-mode regime for the transverse
orientation of dipoles, so that the dispersion branch becomes
backward for magnetic scatterers, and forward for electric ones.
\end{itemize}

Finally, we would like to emphasize, that the width of the pass band for the waveguide loaded by 
transversal electric scatterers has the same order as in the case of transversal magnetic scatterers 
(Marques waveguide \cite{Marqueswaveguide,Hrabarwaveguide}) if the loading scatterers have parameters
obtained using duality principle from each other.
So, the loading by electric scatterers could be an alternative and even more appropriate solution for the
waveguide miniaturization than the design suggested in \cite{Hrabarwaveguide} for this purpose.

\section{Conclusion}

The dispersion properties of rectangular waveguides loaded by
resonant scatterers (magnetic and electric ones) have been
studied. The waveguide problem has been transformed using the
image theory into the eigenmode problem of an auxiliary
three-dimensional electromagnetic crystal. The dispersion
properties of such electromagnetic crystal have been modelled
using the local field approach. It has been revealed that not only
magnetic but also electric resonant scatterers allow to obtain 
mini pass band below cutoff frequency of the hollow waveguide.
The corresponding mini-band turns out to be
of the same order as for magnetic scatterers with same individual frequency
dispersion. So, the electric scatterers (inductively loaded short
wires) could be also prospective for the waveguide miniaturization as well as split-ring-resonators.
It has been shown that the loading by scatterers with longitudinal orientation of dipole moments
also allows to obtain the mini-band of propagation below the cutoff frequency of the hollow waveguide,
but the width of this band is significnatly narrower as compared to the case of transverse orientation.
The observed effects are explained in terms of the subwavelength guiding properties of the single chains of scatterers.
This explanation is supported by comparison of dispersion properties of the loaded waveguides
and the chains of the resonant scatterers in free space.
Results of our theory are in good agreement with the known literature data. 
For the chains of resonant dipoles the results from \cite{Weber} are reproduced. 
For the rectangular waveguide loaded by split-ring-resonators
the same result as in \cite{Marqueswaveguide} has been obtained.

\bibliography{MW}

\appendix

\section{Polarizabilities of resonant scatterers}

\subsection{Split-Ring Resonators}

The SRR considered in \cite{PendrySRR,SmithWSRR,Shelbyscience} is
a pair of two coplanar broken metal rings (see Fig.\ref{scat}.b).
Since the two loops of an SRR are not identical the analytical
models of it are rather cumbersome \cite{MarquesSRR,SimSRR}. In
fact, such SRR can not be described as a purely magnetic
scatterer, because it exhibits bianisotropic properties and has
resonant electric polarizability \cite{MarquesSRR,SimSRR} (see
also discussion in \cite{APSWSRR}). However, the electric
polarizability and bianisotropy of SRR is out of the scope of this
paper. We neglect these effects and consider an ordinary SRR as a
magnetic scatterer. The analytical expressions for the magnetic
polarizability $\alpha(\omega)$ of SRRs with geometry plotted in
Fig.\ref{scat}.a were derived and validated in \cite{SimSRR}. The
final result reads as follows: \e
\alpha(\omega)=\frac{A\omega^2}{\omega_0^2-\omega^2+j\omega\Gamma},
\qquad A=\frac{\mu_0^2\pi^2r^4}{L+M}, \l{alpha} \f where
$\omega_0$ is the resonant frequency of magnetic polarizability:
$$
\omega_0^2=\frac{1}{(L+M)C_r},
$$
$L$ is inductance of the ring (we assume that both rings have the
same inductance):
$$
L=\mu_0 r\left[\log\left(\frac{32R}{w}\right)-2\right],
$$
$M$ is mutual inductance of the two rings:
$$
M=\mu_0 r\left[(1-\xi)\log\left(\frac{4}{\xi}\right)-2+\xi\right],
\qquad \xi=\frac{w+d}{2r},
$$
$C_r$ is the effective capacitance of the SRR:
$$
C_r=\varepsilon_0 \frac{r}{\pi} {\rm arccosh}
\left(\frac{2w}{d}\right),
$$
$\Gamma$ is the radiation reaction factor:
$$
\Gamma=\frac{A\omega k^3}{6\pi\mu_0},
$$
$r$ is the inner radius of the inner ring, $w$ is the width of the
rings, $d$ is distance between the edges of the rings (see
Fig.\ref{scat}.a), $\varepsilon_0$ and $\mu_0$ are permittivity and
permeability of the host media, and
$k=\omega\sqrt{\varepsilon_0\mu_0}$ is the wave number of the host
medium. The presented formulae are valid within the frame of the
following approximations: $w,d\ll r$ and the splits of the rings
are large enough compared to $d$. Also, we assume that SRR is
formed by ideally conducting rings (no  dissipation losses).

The magnetic polarizability \r{alpha} takes into account the
radiation losses and satisfies to the basic Sipe-Kranendonk
condition \cite{Sipe,Belovcond,Belovnonres} which in the present
case has the following form: 
\e 
{\rm Im}\left\{\alpha^{-1}(\omega)\right\}=\frac{k^3}{6\pi\mu_0}.
\l{sipe} \f

In our analysis we operate with the inverse
polarizability $\alpha^{-1}(\omega)$, thus, we rewrite \r{alpha}
in the following form: \e
\alpha^{-1}(\omega)=A^{-1}\left(\frac{\omega_0^2}{\omega^2}-1\right)+j\frac{k^3}{6\pi\mu_0}.
\l{invalph} \f

\subsection{Inductively Loaded Short Wires}

An inductively loaded short wire is shown in Fig. \ref{scat}.b.
The electric polarizability $\alpha_e$ of an inductively loaded
wire following the known model \cite{LWD} has the form: \e
\alpha_e^{-1}= \frac{3}{l^2 C_{\rm wire}}
\left(\frac{1-\omega^2/\omega_0^2}{4-\omega^2/\omega_0^2}\right)+
j\frac{k^3}{6\pi\varepsilon_0} \l{alpe} \f where
$C_{\mbox{wire}}=\pi l\varepsilon_0/\log(2l/r_0)$ is the
capacitance of the wire, $\omega_0=\sqrt{L C_{\rm wire}}$ is the
resonant frequency,  $L$ is the inductance of the load, $l$ is the
half length of the wire and $r_0$ is the wire radius.

It is clear, that at the frequencies near the resonance the
polarizability of LSW has the form \e \alpha_e^{-1}(\omega)\approx
A_e^{-1}\left(\frac{\omega_0^2}{\omega^2}-1\right)+j\frac{k^3}{6\pi\varepsilon_0},
\l{invalphe} \f with $A_e=l^2 C_{\rm wire}$, which is similar to
\r{invalph}. Moreover, if $A_e/\varepsilon_0=A/\mu_0$ then using
duality principle the magnetic dipole with polarizability $\alpha$
\r{invalph} can be transformed to the electric dipole with
polarizability $\alpha_e$ \r{invalph}, and vice versa. This means
that it is enough to consider only one type of resonant
scatterers. In the present paper we have chosen magnetic ones to
be principal. The case of electric scatterers was obtained using
the duality principle with $A=\mu_0A_e/\varepsilon_0$.

\section{Interaction constants of the chains}

The initial expressions for the interaction constants $C_{x,y}$ entering
\r{displine} follow from \r{Green} and \r{gre} and read as follows:
\e 
C_x=\sum\limits_{m\ne 0}\frac{1+jka|m|}{2\pi
a^3|m|^3}e^{-j(k|m|+qm)a} \l{Cx} 
\f
$$
=\frac{1}{\pi a^3}\sum\limits_{m=1}^{+\infty}
\left[\frac{1}{m^3}+\frac{jka}{m^2}\right]e^{-jkam}\cos(qam),
$$
for the longitudinal polarization, and
\e C_y=\sum\limits_{m\ne 0} \frac{k^2a^2m^2-jka|m|-1}{4\pi
a^3|m|^3}e^{-j(k|m|+qm)a} \l{Cy} \f
$$
=\frac{1}{2\pi a^3}\sum\limits_{m=1}^{+\infty}
\left[\frac{k^2a^2}{m}-\frac{jka}{m^2}-\frac{1}{m^3}\right]e^{-jkam}\cos(qam)
$$
for the transverse one (see e.g. \cite{Weber}).

Note, that $C_x$ includes only near-field terms (of the order
$1/R^2$ and $1/R^3$). In contrast to $C_x$, the transverse
interaction constant $C_y$ includes also the wave terms (of the order $1/R$) which
corresponds to the slowly converging series. It makes the direct
numerical summation of \r{Cy} to be not efficient. The series in
\r{Cx} have better convergence, but it is also not enough for rapid calculations.

The application of acceleration technique done in \cite{Belovhomo}
offers a more cumbersome expression for $C_x(k,q,a)$ than \r{Cx},
but it is better for numerical calculations since the series
converges very rapidly:
$$
C_x=\frac{1}{4\pi a^3} \left[ 4\sum\limits_{m=1}^{+\infty}
\frac{(2jka+3)m+2}{m^3(m+1)(m+2)}e^{-jkam}\cos(qam) \right.
$$
\e -(jka+1)\left(t_+^2\log t^++t_-^2\log t^-+2e^{jka}\cos
(qa)\right) \l{c3} \f
$$
\left. -2jka\left(t_+\log t^++t_-\log t^-\right) +(7jka+3)
\vphantom{\sum\limits_{m=1}^{+\infty}
\frac{(2jka+3)m+2}{m^3(m+1)(m+2)}} \right],
$$
where
$$
t^+=1-e^{-j(k+q)a}, \qquad t^-=1-e^{-j(k-q)a},
$$
$$
t_+=1-e^{j(k+q)a}, \qquad t_-=1-e^{j(k-q)a}.
$$

As to $C_y$, the series of the order $1/m$ in \r{Cy} can be
obtained in the closed form using the tabulated formula (see
\cite{Collin}, Appendix): \e \sum\limits_{m=1}^{+\infty} \frac{e^{
-j\gamma m}}{m}= -\log \left(1-e^{-j\gamma}\right) \l{sumn} \f
$$=-\left(\log \left|2\sin \frac{s}{2}\right|+j\frac{\pi
-\gamma'}{2}\right),
$$ where $\gamma'=2\pi\{\gamma/(2\pi)\}$ and we
use notation $\{x\}$  for fractional part of variable $x$. The
rest part of the expression \r{Cy} is simply proportional to
$C_x$. Thus, $C_y$ can be evaluated as follows:
$$
C_y(k,q,a)=-\frac{k^2}{4\pi a} \log \left|2 \left( \cos ka- \cos
qa \right)\right|
$$
\e -
j\frac{k^2}{4a}\left(1-\left\{\frac{(k+q)a}{2\pi}\right\}-\left\{\frac{(k-q)a}{2\pi}\right\}\right)
-C_x(k,q,a)/2. \l{Cyx} \f

In the works \cite{Belovhomo,Tretlines} it was shown that \e {\rm
Im} (C_x)=\frac{k^3}{6\pi}+\frac{1}{4a}\sum\limits_{|q_m|<k}
\left(q_m^2-k^2\right), \l{imcx} \f where
$$
q_m=q+\frac{2\pi m}{a}.
$$

From the formula \r{Cy} using some algebra  it follows that \e
{\rm Im} (C_y)=\frac{k^3}{6\pi}-\frac{1}{8a}\sum\limits_{|q_m|<k}
\left(q_m^2+k^2\right). \l{imcy} \f

The expressions \r{imcx} and \r{imcy} demonstrate energy
transformations happening in the chains. A single scatterer
radiates cylindrical wave and that is why its polarizability has
radiation losses which can be described by Sipe-Kronendonk
condition \r{sipe}. Being arranged into the regular chains the
scatterers acquire effective polarizability (with respect to the external field) of the form: 
\e
\alpha_{x,y}=\left[\alpha^{-1}-\mu_0^{-1}C_{x,y}\right]^{-1}, 
\f
where indices $x$ and $y$ correspond to longitudinal and
transverse orientations of scatterers in the chain, respectively.
In accordance to \r{imcx} and \r{imcy} one can formulate the following
analogues of Sipe-Kronendonk condition for effective
polarizabilities of the scatterers in the chains: \e {\rm Im}
\left\{\alpha_x^{-1}\right\}=\frac{1}{4\mu_0a}\sum\limits_{|q_m|<k}
\left(k^2-q_m^2\right), \l{imax} \f \e {\rm Im}
\left\{\alpha_y^{-1}\right\}=\frac{1}{8\mu_0a}\sum\limits_{|q_m|<k}
\left(q_m^2+k^2\right). \l{imay} \f

Note, that terms $k^3/(6\pi)$ are canceled. It is clear, that if
there are no such index $m$ that $|q_m|<k$ (like it happens for
example if $k<q<2\pi/a-k$ for $k<\pi/a$) then effective
polarizabilities turn out to be purely real and the chain itself
does not radiate. This regime corresponds to the case of guiding
modes and it makes dispersion equation \r{displine} real valued
one. If there are some indices $m$ that $|q_m|<k$ then the
effective polarizabilites acquire non-zero imaginary part which
give evidence that the chain radiates cylindrical waves. The
number of such waves corresponds to the number of indices $m$
fulfilling to the relation $|q_m|<k$. If $|q|<k<\pi/a$ then the
chain radiates only one cylindrical wave which can be treated as
the main diffraction lobe of this periodical array. For the
higher frequencies the grating lobes appears and all of them make
its contribution to expressions \r{imax} and \r{imay}.

\section{Interaction constant of an orthorhombic lattice}

For effective numerical calculation of interaction constant $C(k,\-q,a,b,c)$ defined by \r{C} 
we use the following formula deducted in \cite{Belovhomo}:
$$
C(k,\-q,a,b,c)= -\sum\limits_{n=1}^{+\infty}\sum\limits_{{\rm
Re}(p_m)\ne 0} \frac{p_m^2}{\pi a}K_0\left(p_mbn\right)\cos(q_ybn)
$$
\e
+\sum\limits_{m=-\infty}^{+\infty}\sum\limits_{n=-\infty}^{+\infty}
\frac{p_m^2}{2jab k_z^{(mn)}} \frac{e^{-j k_z^{(mn)}c}-\cos
q_zc}{\cos k_z^{(mn)}c-\cos q_zc} \l{cfinal} \f
$$
-\sum\limits_{{\rm Re}(p_m)=0} \frac{p_m^2}{2ab}
\left(\frac{1}{jk_z^{(m0)}}+ \sum\limits_{n=1}^{+\infty}
\left[\frac{1}{jk_z^{(m,n)}}+\frac{1}{jk_z^{(m,-n)}}\right.
\right.
$$
$$
\left.\left. -\frac{b}{\pi n}-\frac{l_mb^3}{8\pi^3
n^3}\right]+1.202 \frac{l_mb^3}{8\pi^3}+ \frac{b}{\pi} \left(\log
\frac{b|p_m|}{4\pi}+\gamma\right)+j\frac{b}{2}
\vphantom{\sum\limits_{n\ne 0}
\left[\frac{1}{jk_z^{(mn)}}-\frac{b}{2\pi |n|}\right]}\right)
$$
$$
+C_x(k,q_x,a),
$$
where $C_x$ is given by \r{c3} and
$$
k_x^{(m)}=q_x+\frac{2\pi m}{a},\qquad k_y^{(n)}=q_y+\frac{2\pi
n}{b},
$$
$$
p_m=\sqrt{\left(k_x^{(m)}\right)^2-k^2},\qquad l_m=2q_y^2-p_m^2,
$$
$$
k_z^{(mn)}=-j\sqrt{\left(k_x^{(m)}\right)^2+\left(k_y^{(n)}\right)^2-k^2}.
$$
\end{document}